\documentclass[sigplan,9pt,screen]{acmart}

\usepackage{breqn}
\usepackage{filecontents}
\usepackage{algorithm}
\usepackage{algpseudocode}
\usepackage{multirow}
\usepackage{makecell}
\usepackage{fancyhdr}
\usepackage{listings}
\usepackage{balance}
\usepackage{amsmath,amsfonts}
\usepackage{graphicx}
\usepackage{epstopdf}
\usepackage{textcomp}
\usepackage{xcolor}
\usepackage{subcaption} % 在导言区加载包
\usepackage{float}%提供float浮动环境
\usepackage{booktabs}%提供命令\toprule、\midrule、\bottomrule
\usepackage{algorithm}
\usepackage{algpseudocode}
\usepackage{setspace}
\usepackage{svg} % 需使用包
\usepackage{etoolbox}
\usepackage{url}
\usepackage{newtxtext,newtxmath}
\usepackage{tikz}
\usepackage{tabularx}
\usepackage{booktabs}
\usepackage{multirow}
\usepackage{makecell}
\usepackage{array}
\usepackage{cancel}
\usepackage{colortbl}
\usepackage{algpseudocode}
\usepackage{graphicx}
\usepackage[font=small,labelfont=bf,textfont=bf]{caption}
\usepackage{fancyhdr}
\usepackage{balance}

\usepackage[normalem]{ulem}

\newrobustcmd*\blacka[1]{\tikz[baseline=(char.base)]{
            \node[shape=circle,draw,inner sep=1pt,fill,text=white,minimum size=1em] (char) {\textsf{\small a}};}}

\newrobustcmd*\blackb[1]{\tikz[baseline=(char.base)]{
            \node[shape=circle,draw,inner sep=1pt,fill,text=white,minimum size=1em] (char) {\textsf{\small b}};}}

\newrobustcmd*\blackc[1]{\tikz[baseline=(char.base)]{
            \node[shape=circle,draw,inner sep=1pt,fill,text=white,minimum size=1em] (char) {\textsf{\small c}};}}

\newrobustcmd*\blackd[1]{\tikz[baseline=(char.base)]{
            \node[shape=circle,draw,inner sep=1pt,fill,text=white,minimum size=1em] (char) {\textsf{\small d}};}}

\newrobustcmd*\blacke[1]{\tikz[baseline=(char.base)]{
            \node[shape=circle,draw,inner sep=1pt,fill,text=white,minimum size=1em] (char) {\textsf{\small e}};}}

\newrobustcmd*\blackf[1]{\tikz[baseline=(char.base)]{
            \node[shape=circle,draw,inner sep=1pt,fill,text=white,minimum size=1em] (char) {\textsf{\small f}};}}

\newrobustcmd*\blackg[1]{\tikz[baseline=(char.base)]{
            \node[shape=circle,draw,inner sep=1pt,fill,text=white,minimum size=1em] (char) {\textsf{\small g}};}}

\newrobustcmd*\blackh[1]{\tikz[baseline=(char.base)]{
            \node[shape=circle,draw,inner sep=1pt,fill,text=white,minimum size=1em] (char) {\textsf{\small h}};}}

\newrobustcmd*\blacki[1]{\tikz[baseline=(char.base)]{
            \node[shape=circle,draw,inner sep=1pt,fill,text=white,minimum size=1em] (char) {\textsf{\small i}};}}
            
\newrobustcmd*\blackj[1]{\tikz[baseline=(char.base)]{
            \node[shape=circle,draw,inner sep=1pt,fill,text=white,minimum size=1em] (char) {\textsf{\small j}};}}
            
\newrobustcmd*\blackk[1]{\tikz[baseline=(char.base)]{
            \node[shape=circle,draw,inner sep=1pt,fill,text=white,minimum size=1em] (char) {\textsf{\small k}};}}

\newrobustcmd*\blackasmall[1]{\tikz[baseline=(char.base)]{
  \node[shape=circle, draw, inner sep=0.8pt, fill, text=white, minimum size=0.8em] (char) {\textsf{\scriptsize a}};}}
  
\newrobustcmd*\blackbsmall[1]{\tikz[baseline=(char.base)]{
  \node[shape=circle, draw, inner sep=0.7pt, fill, text=white, minimum size=0.7em] (char) {\textsf{\scriptsize b}};}}

  \newrobustcmd*\blackcsmall[1]{\tikz[baseline=(char.base)]{
  \node[shape=circle, draw, inner sep=0.8pt, fill, text=white, minimum size=0.8em] (char) {\textsf{\scriptsize c}};}}

  \newrobustcmd*\blackdsmall[1]{\tikz[baseline=(char.base)]{
  \node[shape=circle, draw, inner sep=0.8pt, fill, text=white, minimum size=0.8em] (char) {\textsf{\scriptsize d}};}}

  \newrobustcmd*\blackesmall[1]{\tikz[baseline=(char.base)]{
  \node[shape=circle, draw, inner sep=0.7pt, fill, text=white, minimum size=0.7em] (char) {\textsf{\scriptsize e}};}}

  \newrobustcmd*\blackfsmall[1]{\tikz[baseline=(char.base)]{
  \node[shape=circle, draw, inner sep=0.7pt, fill, text=white, minimum size=0.7em] (char) {\textsf{\scriptsize f}};}}

  \newrobustcmd*\blackgsmall[1]{\tikz[baseline=(char.base)]{
  \node[shape=circle, draw, inner sep=0.7pt, fill, text=white, minimum size=0.7em] (char) {\textsf{\scriptsize g}};}}
  
  \newrobustcmd*\blackhsmall[1]{\tikz[baseline=(char.base)]{
  \node[shape=circle, draw, inner sep=0.8pt, fill, text=white, minimum size=0.8em] (char) {\textsf{\scriptsize h}};}}

  \newrobustcmd*\blackismall[1]{\tikz[baseline=(char.base)]{
  \node[shape=circle, draw, inner sep=0.8pt, fill, text=white, minimum size=0.8em] (char) {\textsf{\scriptsize i}};}}

  \newrobustcmd*\blackjsmall[1]{\tikz[baseline=(char.base)]{
  \node[shape=circle, draw, inner sep=0.8pt, fill, text=white, minimum size=0.8em] (char) {\textsf{\scriptsize j}};}}

    \newrobustcmd*\blackksmall[1]{\tikz[baseline=(char.base)]{
  \node[shape=circle, draw, inner sep=0.7pt, fill, text=white, minimum size=0.7em] (char) {\textsf{\scriptsize k}};}}

\setcopyright{none}
\begin{document}

\title{Strix: Re-thinking NPU Reliability from a System Perspective}

\newcommand{\zhenote}[1]{\textbf{\textcolor{blue}{Zhe: #1}}}
\newcommand{\guan}[1]{\textbf{\textcolor{red}{Guan: #1}}}

\newcommand{\npumodel}{Strix}

\makeatletter
\setlength{\footnotesep}{4pt}    % 多条脚注之间的距离
\makeatother

\author{Jiapeng Guan$^{\dagger}$, Jie Zhang$^\S$, Hao Zhou$^{\dagger}$, Ran Wei$^\ddagger$, Dean You$^\S$, Hui Wang$^\S$, Yingquan Wang$^{\dagger}$, Tinglue Wang$^\S$, Xudong Zhao$^{\dagger}$, Jing Li$^{\P}$ and Zhe Jiang$^\S$$^*$}
\thanks{$^*$Corresponding author:
Zhe Jiang. Email: zhejiang.uk@gmail.com.}

\affiliation{%
  \institution{\hspace*{-1.5em}\mbox{$^{\dagger}$Dalian University of Technology, China. $^\S$Southeast University, China.  $^\ddagger$Lancaster University, UK. $^{\P}$New Jersey Institute of Technology, US.}}
  \country{}
}

\renewcommand{\shortauthors}{Jiapeng Guan et al.}

\begin{abstract}
DNNs and LLMs increasingly rely on hardware accelerators, including in safety-critical domains, while technology scaling and growing model complexity make hardware faults more frequent. 
Existing system-level mechanisms typically treat the NPU as a monolithic unit, using coarse-grained replication that incurs prohibitive performance and hardware overheads, leaving a gap between reliability requirements and deployable solutions.
To bridge this gap, we present Strix, a full-stack NPU reliability framework on an open-source SoC, spanning micro-architecture, ISA, and programming methods. Strix re-partitions the NPU along the system inference pipeline, identifies dominant failure modes, and attaches targeted safeguards, achieving sub-micro-second fault localisation, error detection, and correction with only 1.04$\times$ slowdown and minimal hardware overhead.
% DNNs and LLMs rely on hardware accelerators to satisfy their computational demands, with such accelerators now being deployed in safety-critical domains.
% However, with the transistor scaling, reduced operating voltages, and growing model complexity, hardware faults are becoming more common, imposing a significant barrier to guarantee system reliability.
% Existing mechanisms (e.g., modular or instruction redundancy) typically treat the NPU as a monolithic unit, performing coarse-grained analysis and replication, which incurs substantial performance and hardware overheads.
% Under tight latency, throughput, and hardware budgets, especially in safety-critical systems, such redundancy is often impractical, leaving a gap between reliability requirements and deployable solutions.

% To bridge this gap, we present Strix, a full-stack framework of NPU reliability, on an open-source SoC, from micro-architecture and ISA to the programming methods.
% Strix tackles the above challenge by re-partitioning the NPU along the system inference pipeline, identifying dominant failure modes, and attaching targeted safeguards to each partition, thereby enabling sub-microsecond-level protection capacity (fault localisation, error detection, and correction).
% Through pipeline design and offline configuration, Strix only causes a 1.04$\times$ geometric slowdown, with only 8.7\% area and 16.8\% power overhead, aligning high-coverage protection with practical deployment constraints.
\end{abstract}

\maketitle

\section{Introduction}
\label{sc:Intro}
DNNs and LLMs deliver SOTA accuracy but at steep computational cost~\cite{al2022review}. Modern models move billions to trillions of parameters and perform orders-of-magnitude more Multiply-and-Accumulate (MAC) operations per inference~\cite{kenton2019bert,guo2024deepseek}, pushing SoCs to dedicate substantial area to accelerators (e.g., NPUs and GPUs)~\cite{korosec2021tesla,ditty2022nvidia}. In safety-critical deployments (e.g., autonomous driving~\cite{li2025understanding,fu2024drive}) and with so much hardware in the loop, reliability must be re-examined~\cite{cui2024survey,mittal2020survey}. 
Intrinsic fault tolerance alone is insufficient~\cite{mittal2020survey}; for instance, \cite{li2025understanding} reports a case where a truck was misclassified as a bird due to faults in the NPU.
Such failures are becoming more likely as ever-larger models widen the temporal/spatial exposure to bit-level faults~\cite{kenton2019bert,guo2024deepseek}, device scaling increases vulnerability to variation, radiation, and ageing~\cite{cao2023future}, and intensive NPU data-reuse lets a single error contaminate many activations across layers~\cite{genc2021gemmini,chen2019eyeriss}.

\noindent \textbf{Existing work.} 
Research on NPU reliability can be organised across different architectural levels.
% : algorithmic, instruction, micro-architectural, and system. 
At the algorithmic level, common approaches include injecting faults during training to enhance neuron robustness~\cite{zhou2018adaptive,fromm2018heterogeneous}, employing voting schemes to mitigate the impact of bit flips~\cite{guilleme2024htag}, or modifying activation functions to suppress error propagation~\cite{hoang2020ft,breier2018practical}. 
% However, such methods often struggle to satisfy safety standards (e.g., ISO-26262~\cite{iso201126262}) as they do not provably guarantee timely fault detection and/or correction, and their cross-model scalability remains contested. 
However, such methods do not provide direct error detection, and their cross-model scalability remains contested. 
At the instruction level, researchers rely on Instruction (or code) Redundancy (IR) to protect sensitive layers~\cite{ibrahim2020analyzing,wei2020analyzing,ibrahim2020soft2,saikawa2024approximated}; yet they typically demand layer-by-layer analysis and manual code modifications, and incur substantial performance overhead owing to redundant execution.
At the micro-architectural level, most studies focus on a single component, such as Processing Element (PE)~\cite{taheri2024exploration,lu2024highly,jian2024perft,xie2025realm}, or a single fault model (e.g., transient faults~\cite{li2025understanding,pandey2019greentpu}). 
This localised perspective hinders a unified characterisation of system-level dataflow and timing coupling, overlooking interaction effects among local mechanisms in buffering and timing, and ultimately falling short of end-to-end reliability guarantees.
Moreover, the above approaches fail to meet the system-wide error detection requirements mandated by safety standards (e.g., ISO-26262~\cite{iso201126262}).
% \zhenote{Can we make an argument: algorithmic, instruction, and micro-architectural protection can not provide enough protection? The expected protection HAVE TO BE considered at the system level. 
% However, the system-level protection can only use TMR.}
Achieving system-level protection still relies on costly triple modular redundancy (TMR)~\cite{libano2018selective,sanchez2016error,wei2023approximate}.
% Although it offers comprehensive coverage of NPU reliability, its overhead renders it impractical for many real deployments, leaving a persistent gap between reliability requirements and deployable solutions. \zhenote{Please highlight the analysis and protection is too crosa-grainted -- this is the key reason.}
Although it offers comprehensive coverage, its coarse-grained, whole-system analysis and protection incurs prohibitive overhead, rendering it impractical for many deployments.
% and finally leaving a gap between reliability requirements and deployable solutions.
% \zhenote{Here is not enough, we need to tell ``system-level'' safeguards can provide comprehensive protection because it has the global view, but it is too expensive due to the triple modular redudancy.}
% \zhenote{A key here is that, on one hand, we have no choice but that only the system-level scheme can have ASIL-D protection; on the other hand, system-level scheme can only use TMR or DMR, which is very costly --  but is the TMR microarchitectural scheme?}

\noindent \textbf{Research challenges.}
% From a system perspective, we can place the model, hardware, and runtime within a single optimisation space with unified metrics and constraints, enabling end-to-end verifiable reliability under tight resource budgets.
Therefore, to truly achieve a reliable NPU solution that meets the diagnostic requirements of safety standards and provides end-to-end verifiable reliability, we must adopt a system-level perspective, integrating the model, hardware, and runtime into a unified optimisation space.
However, it is challenging due to cross-layer couplings and resource constraints.
A practical, deployable solution must (i) accurately characterise potential failure modes within the system  — avoiding both excessive granularity (e.g., analysing individual ALUs, leading to combinatorial explosion) and overly coarse granularity (e.g., treating the entire NPU as a monolithic unit, which degrades resource utilisation); 
and (ii) implement targeted mitigation strategies; this, in turn, requires a deep understanding of the neural network’s hierarchical structure and data sensitivities, as well as the NPU’s micro-architectural operation.
% (e.g., data reuse and pipelined execution). 
In summary, only by adopting a system view that integrates model behaviour and hardware fault patterns into an analytical framework, and by applying tailored hardening with coordinated orchestration across key subsystems, can we achieve high-coverage, low-slowdown guarantees that meet safety standards within strict budgets.

\begin{figure*}[!t]
\captionsetup[subfigure]{skip=1.5pt}  % 控制图与caption之间的垂直距离
% \centering
% 第一行
\begin{subfigure}[t]{0.31\textwidth}
    \centering
    \includegraphics[width=5.5cm,height=1.95cm]{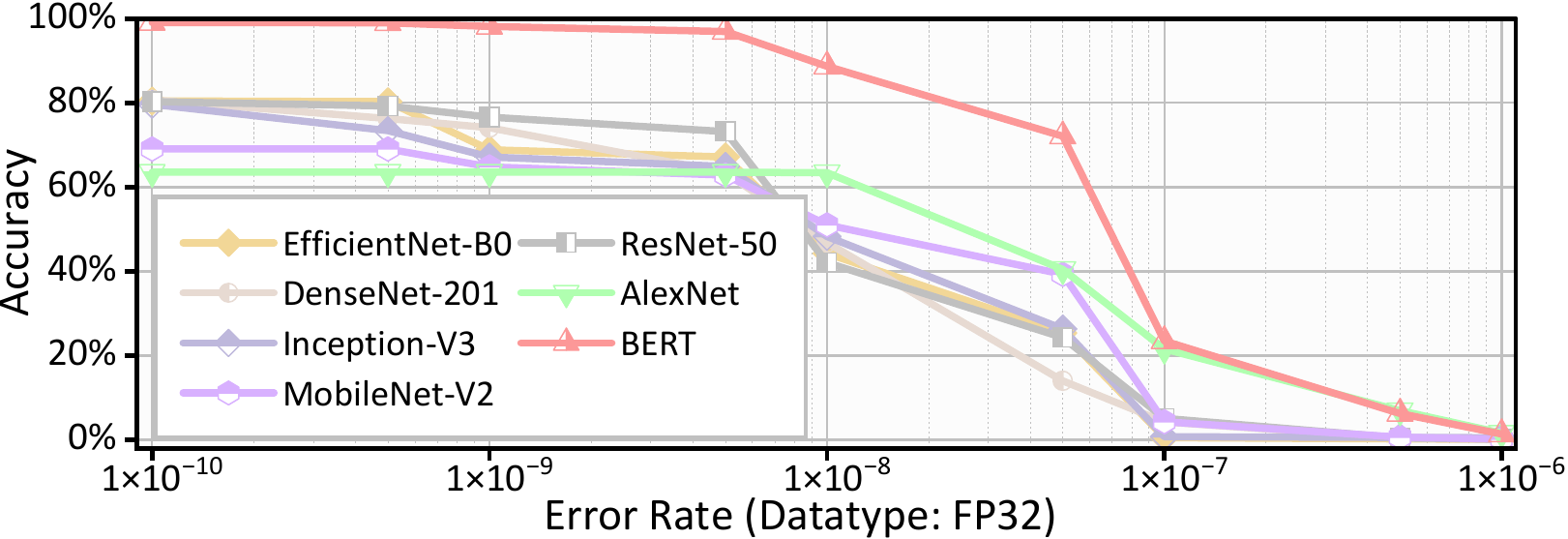}
    \caption{FP32 DNNs under different error rates.}
    \label{fig:moti1}
\end{subfigure}
\hspace{13pt}
\begin{subfigure}[t]{0.31\textwidth}
    % \centering
    \includegraphics[width=5.3cm,height=1.95cm]{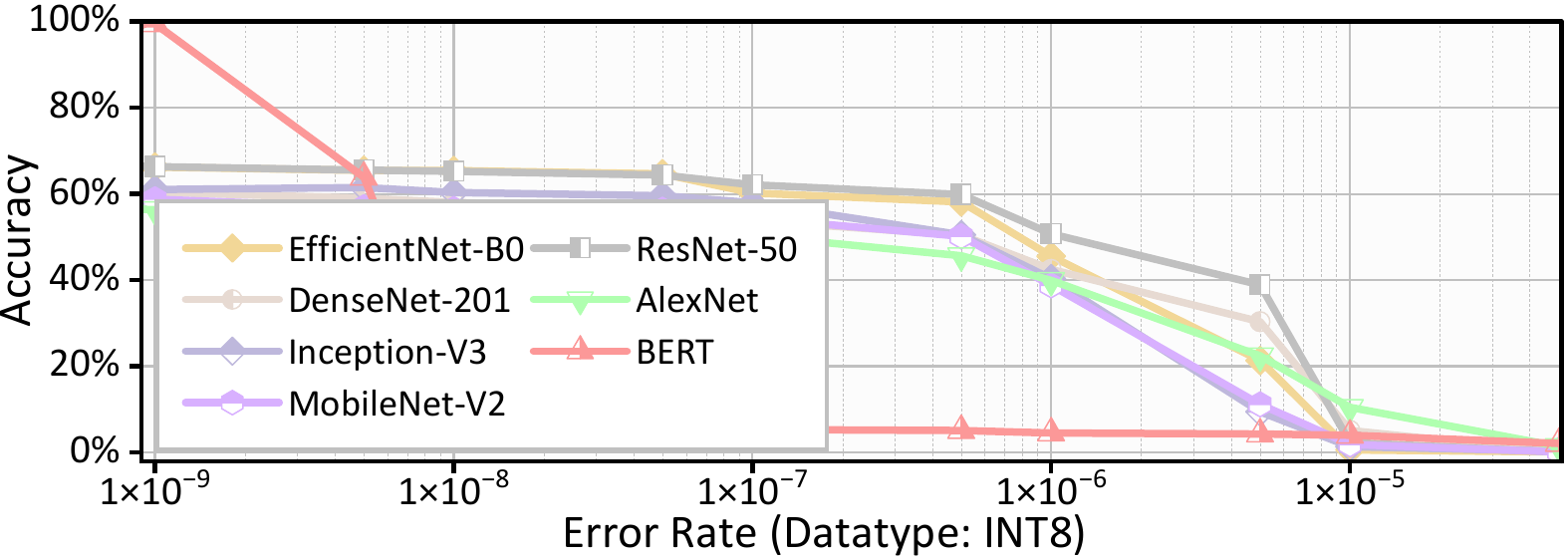}
    \caption{INT8 DNNs under different error rates.}
    \label{fig:moti2}
\end{subfigure}
\hspace{12pt}
\begin{subfigure}[t]{0.31\textwidth}
    % \centering
    \includegraphics[width=5.5cm,height=1.95cm]{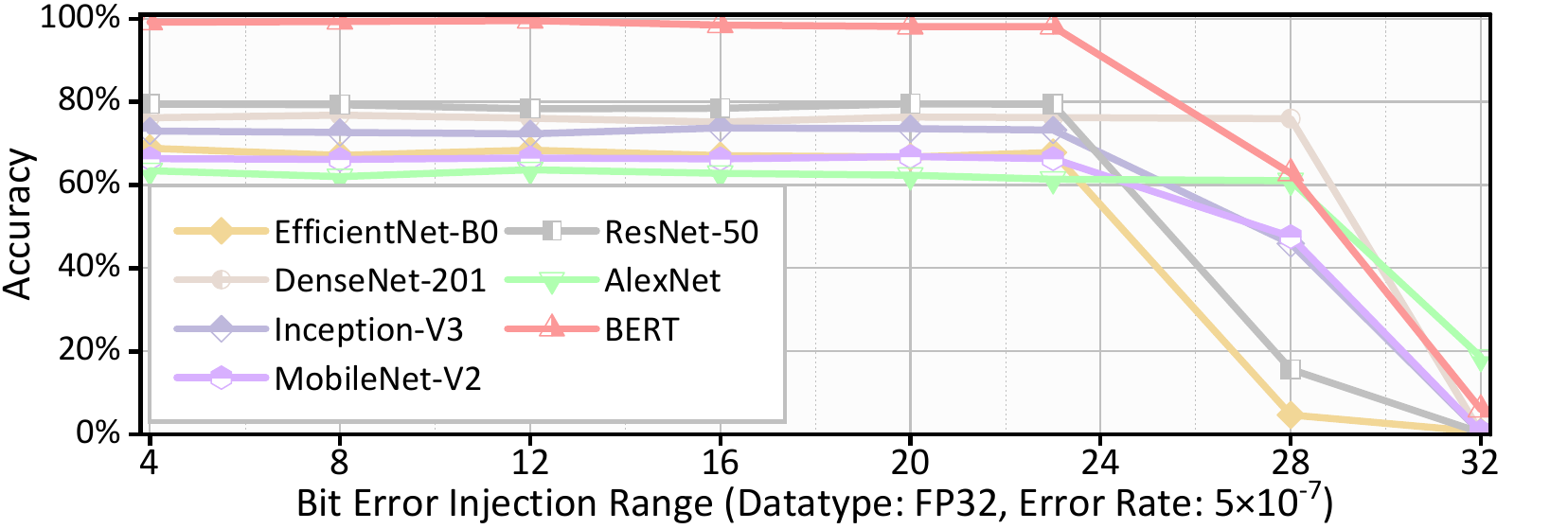}
    \caption{FP32 DNNs under different error positions.}
    \label{fig:moti3}
\end{subfigure}
\hspace{80pt}

% 第二行
\begin{subfigure}[t]{0.31\textwidth}
    \centering
    \includegraphics[height=1.68cm]{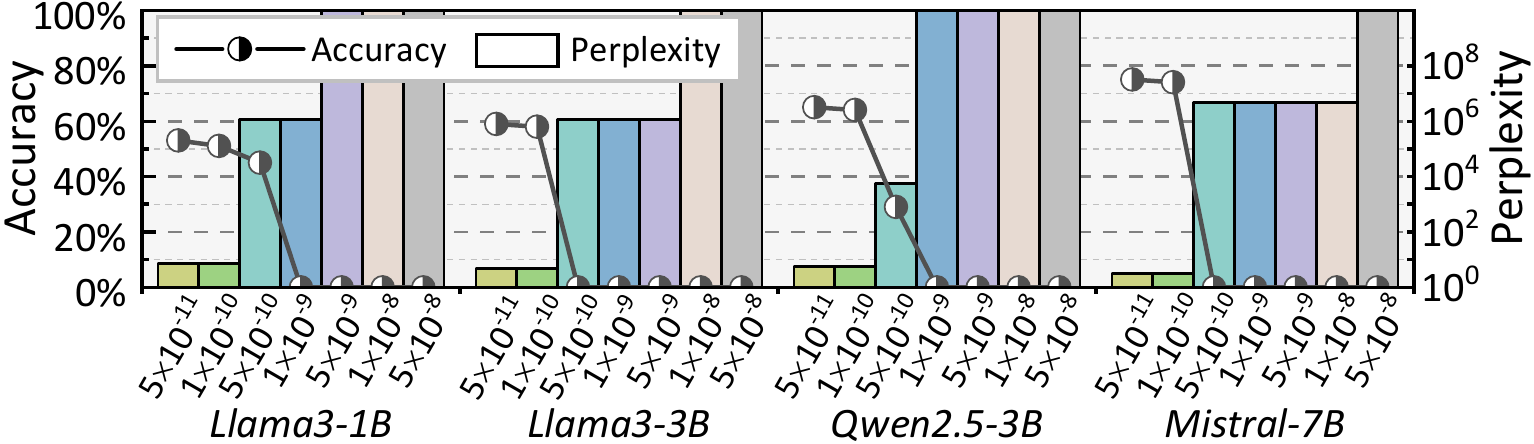}
    \caption{BF16 LLMs under different error rates.}
    \label{fig:moti4}
\end{subfigure}
\hfill
\begin{subfigure}[t]{0.31\textwidth}
    \centering
    \includegraphics[height=1.68cm]{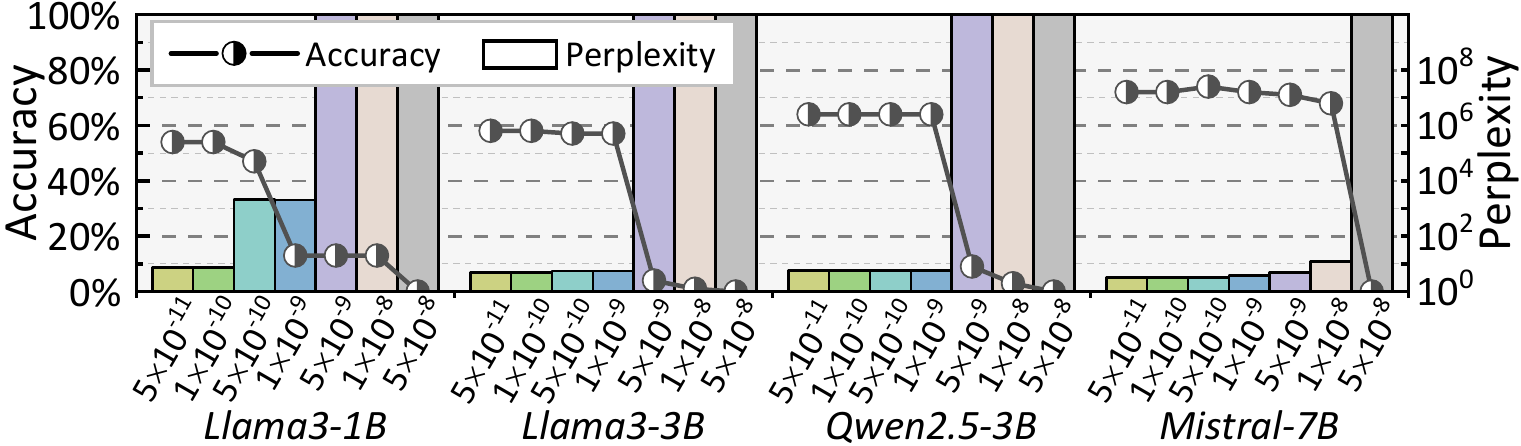}
    \caption{INT8 LLMs under different error rates.}
    \label{fig:moti5}
\end{subfigure}
\hfill
\begin{subfigure}[t]{0.31\textwidth}
    \centering
    \includegraphics[height=1.68cm]{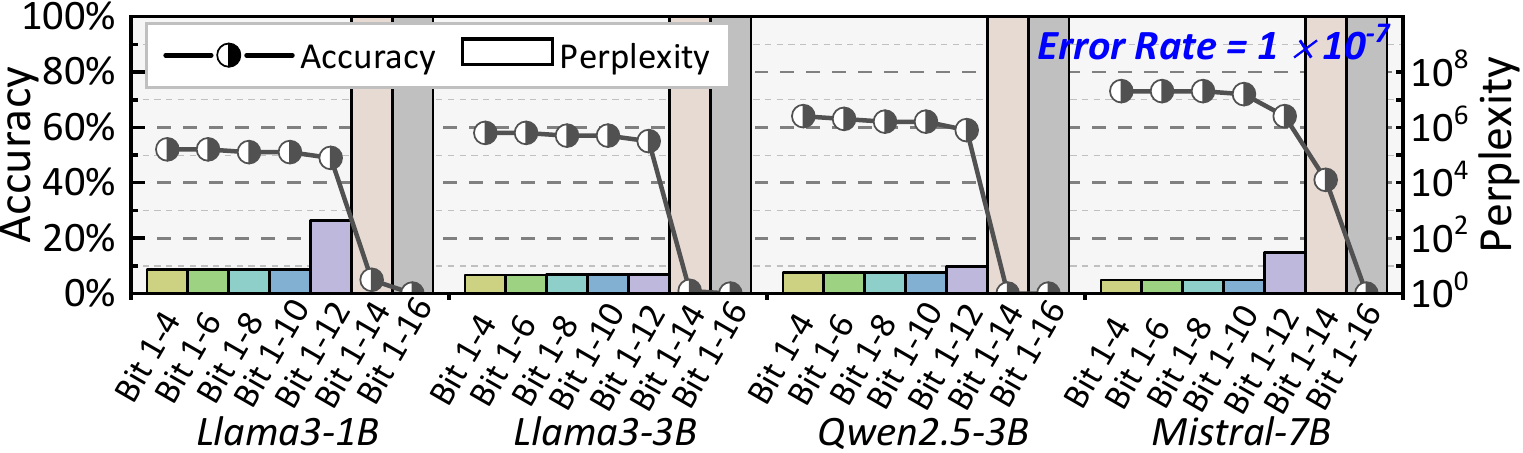}
    \caption{BF16 LLMs under different error positions.}
    \label{fig:moti6}
\end{subfigure}
\vspace{-11pt}
\caption{Performance of different models~\cite{kenton2019bert,tan2019efficientnet,huang2017densely,szegedy2016rethinking,he2016deep,sandler2018mobilenetv2,krizhevsky2012imagenet,dubey2024llama,hui2024qwen2,Mistral7B} under varying error rates and bit positions. The BERT model was tested on the DBPedia-14 dataset~\cite{rangwani2022cost}, while other DNNs were tested on Tiny ImageNet dataset~\cite{yao2015tiny}, and LLMs were evaluated on the LAMBADA dataset~\cite{paperno2016lambada}. 
% Performance degradation under different fault rates highlights the need for reliability support in neural networks, while fault injection across different bit positions reveals that errors in certain positions have a disproportionately large impact on model outcomes.
% \zhenote{Give an summary of it.}
}
\label{fig2}
% \vspace{-12pt}
\end{figure*}

\noindent \textbf{Contributions.}
% \vspace{200pt}
Guided by an ``analyse-first, protect-exactly''  philosophy, we present \textbf{Strix}, a full-stack
% \zhenote{As we discussed, full-stack is reasonable.} 
reliability framework for NPUs.
Strix addresses the heterogeneous failure modes of key components, 
including registers, local memory, systolic array, and non-linear operators, 
through co-designed SoC architecture, ISA extensions, and programming methods.
It establishes hierarchical mechanisms for fault localisation, error detection, and correction, supported by a pipeline and configuration strategies to minimise system overhead.
% \zhenote{Be precise, what types of the theoretical model and optimisation strategies.}
% On a prototype cluster composed of 20 AMD Virtex UltraScale+ VU19P FPGAs,
Experimental results show that Strix maintains a performance loss below $1.07\times$, outperforming TMR ($1.83\times$ to $3.68\times$), IR ($1.13\times$ to $1.95\times$), and ReaLM~\cite{xie2025realm} (a SOTA scheme, $1.22\times$ to $1.59\times$). Under a typical fault rate (i.e., $<10^{-6}$), Strix mitigating performance degradation in neural networks.
In an industrial-scale configuration, Strix incurs only 8.7\% area and 16.8\% power overhead, offering a high-reliability and cost-effective solution for safety-critical applications.

\section{Motivations}
\label{sec:motivation}

% \zhenote{Why do we need to analyse the fault tolerance of the workloads? This is because we attempt to find the vulnerability of the tasks and apply the corresponding safeguards.}
We begin by analysing the fault tolerance of NPU workloads to identify task-level vulnerabilities and guide the placement of safeguards. 
To this end, we retrained several DNN models for classification tasks and selected several pretrained LLMs for language prediction.
We then injected errors using the framework in~\cite{reagen2018ares}. 
Considering the widespread adoption of FP32 and INT8 in DNNs~\cite{dettmers2022gpt3,johnson2018rethinking}, and the compatibility of BF16 and INT8 within LLMs~\cite{ahn2023performance,zhou2024survey}, we evaluated all these representations. 
Finally, guided by an analysis of the NPU pipeline, we partitioned the NPU into four key modules from a reliability perspective and explored the corresponding failure modes.
% Finally, guided by a dataflow analysis of the NPU pipeline, we identified the protection targets as four vulnerability. 
% \zhenote{This wrong sentence. The correct presentation is: we partition the NPU system into four key modules from the reliability perspective, and explore the vulnerability correspondingly.}
% \textbf{\textcolor{red}{I still feel ``vulnerability'' is for security. Can we check the terminologies in the existing work?}}

\noindent \textbf{Impact of error rate.}
As shown in Figs.~\ref{fig2}(\subref{fig:moti1}), \ref{fig2}(\subref{fig:moti2}), \ref{fig2}(\subref{fig:moti4}) and \ref{fig2}(\subref{fig:moti5}), we injected errors into different models, obtaining results consistent with prior observations~\cite{ibrahim2020soft,mittal2020survey}. 
The results validate that machine learning workloads exhibit a degree of fault tolerance. 
For instance, Figs.~\ref{fig2}(\subref{fig:moti1}) and \ref{fig2}(\subref{fig:moti4}) show that, under low error rates (e.g., $\leq 1\times10^{-10}$), the performance remains virtually unchanged. However, when the error rate exceeds a certain threshold, model accuracy drops sharply and perplexity rises rapidly. 
Say, for Llama3‑3B with BF16, the threshold is $5\times10^{-10}$.
% , driving the accuracy to 0 and inflating the perplexity by a factor of $10^{6}$ relative to its baseline. 
In safety-critical systems, silent-error rates in compute-intensive devices exceeding $10^{-7}$ make NPU workloads extremely vulnerable~\cite{gizopoulos2025dark}. 
\textsl{Thus, although NPU workloads exhibit a degree of fault tolerance, this does not enable them to operate accurately in the absence of reliability support.}
% \zhenote{Conclusion??}

\noindent \textbf{Impact of error position.}
As shown in Figs.~\ref{fig2}(\subref{fig:moti3}) and \ref{fig2}(\subref{fig:moti6}), we examined the impact of error injection at different positions. The results reveal that, even at high error rates, errors in the mantissa have virtually no effect on the model performance. 
In contrast, errors in the sign bit and exponent bits lead to a marked performance degradation: all LLMs are highly sensitive to the top four bits of BF16, and most DNNs likewise exhibit strong sensitivity to the top nine bits of FP32. 
\textsl{Thus, fault‑tolerance in NPUs should prioritise the protection of those bits (e.g., sign and exponent bits) that exert greater impact on output, achieving more reliability benefit for a given cost.}
% \zhenote{Conclusion?}

% \zhenote{There is a gap between the insight 1, 2 and 3. 1 and 2 are for models, and 3 is for the NPU. What is the bridge between them? and how?}
% \zhenote{I still feel that there is a mislink between models and the NPU.}

\noindent \textbf{Analysis of system failure modes.}
% \zhenote{Reliability-perspective partitioning and failure mode analysis?}
In NPUs, inference begins with instruction fetch–decode–issue, where registers carry scheduling and configuration; the input matrix is then loaded into local memory. 
Data stream into the systolic array to perform MAC, followed by non-linear operators, and the results are finally written back. 
% to the main memory. 
Along this path, and informed by \cite{ibrahim2020soft,mittal2020survey}, we identify four critical classes: (i) registers, whose faults directly perturb instruction/control semantics, with decode or configuration corruption, leading to program crashes; (ii) local memory, where a single error can widely propagate across parallel computations and subsequent layers, yielding systemic output distortion and task failure; (iii) the systolic array, which executes the core matrix MAC kernels and accounts for over 
70\% of inference time~\cite{dos2023understanding}, so faults here rapidly cascade through later layers and substantially degrade end-to-end accuracy; and (iv) non-linear operators, whose selective amplification/saturation effects shape the visibility of upstream errors and their impact on model outputs. 
% \textsl{Through the analysis of these failure models, we can implement targeted protection mechanisms for the different components of the NPU, thereby developing a comprehensive, full-stack reliability framework.}
\textsl{
By adopting a system view along the NPU pipeline, this failure modes analysis allows us to tailor differentiated reliability strategies to each module, according to its role in the dataflow and its fault characteristics, thereby enabling a comprehensive, full-stack reliability framework that spans the micro-architecture while providing end-to-end protection with minimal overhead.
}

\begin{figure*}[t]
\centering
\includegraphics[width=\linewidth]{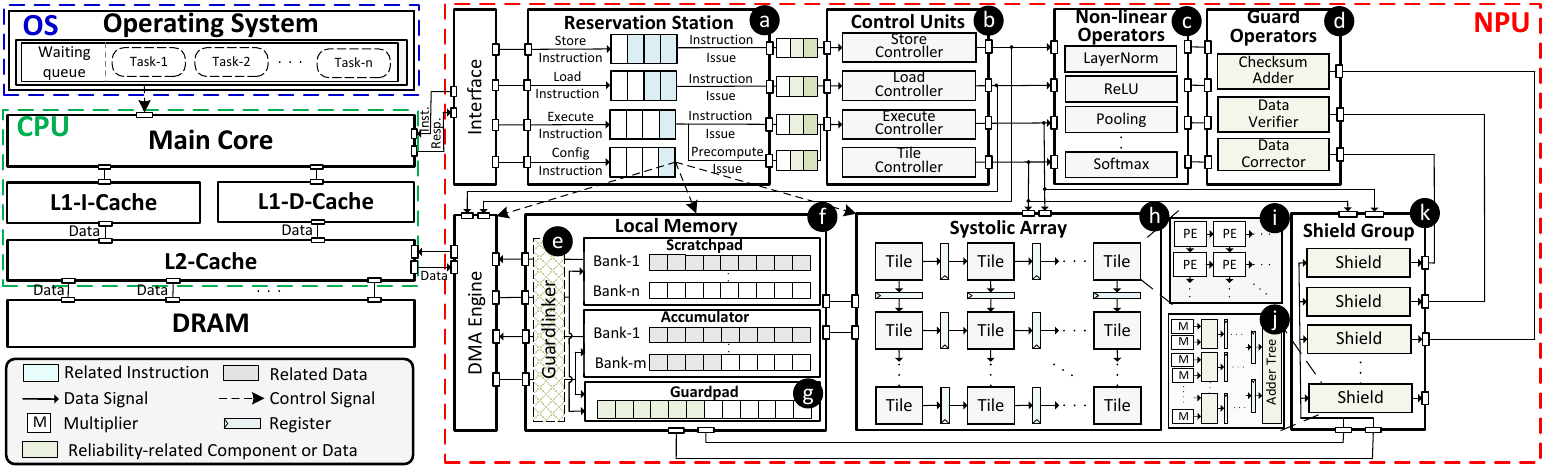}
\vspace{-10pt}
\caption{The architectural overview of Strix.
Strix applies module-specific hardware safeguards for NPUs.
% : SEC-DED ECC for critical registers (\blackasmall\ and \blackbsmall\ ); a guardpad (\blackgsmall\ ) and guardlinker (\blackesmall\ ) in local memory (\blackfsmall\ ) to record and associate integrity metadata; a shield group (\blackksmall\ ) that protects the systolic array (\blackhsmall\ ) in parallel; and lightweight checks for non-linear operators (\blackcsmall\ ) by combining specific properties with TMR.
}
\label{overview}
\vspace{-5pt}
\end{figure*}

 \section{Strix: an Overview}
\label{sec:approach}

To validate Strix, we use the Gemmini~\cite{genc2021gemmini} NPU as a case study.
We choose Gemmini for three main reasons: (i) representative, Gemmini features a representative design similar to other NPUs~\cite{qin2020sigma,shin2018dnpu,chen2019eyeriss,amert2017gpu};
(ii) widely adopted, Gemmini has been widely integrated into SoCs~\cite{gookyi2023deep, vieira2023gem5} and has already undergone tape-out;
(iii) realistic, unlike convolution-centric NPUs~\cite{lian2019high}, Gemmini also supports transformer and integrates non-linear units, making it more aligned with mainstream model acceleration.
% \zhenote{complexity is not the reason. It sounds not like a good thing. How about realistic? support linux etc.?}
The overview is shown in Fig.~\ref{overview}.

\noindent \textbf{System glance.}
% \zhenote{May be it is not the philosophy? It is more like the concept?}
% As analysed in Sec~\ref{sec:motivation},
% instead of protecting the NPU as a whole, we adopt a modular protection strategy.\zhenote{Explicitly explain why do a modular protection strategy.}
Starting from the system pipeline, we identify the key modules and their roles in the dataflow and timing in Sec.~\ref{sec:motivation}. 
% \zhenote{Where? In motivation or Sec.4?}
% Based on this, we adopt a modular protection strategy rather than hardening the NPU as a whole: by tailoring mechanisms to each module and inserting lightweight reliability operators at module boundaries, we provide end-to-end protection with minimal system-level overhead. \zhenote{Different from the coarse-grained, do we also leverage the mathematical characteristics? }
Based on this, we adopt a modular, fine-grained protection strategy rather than hardening the NPU as a whole.
Exploiting the micro-architectural traits of each key module and the mathematical properties of its operators, we tailor safeguards per module and place lightweight reliability operators at their boundaries.
Under this decomposition, all intermediate states and outputs pass through protected modules and boundary checks; once each module provides bounded reliability within its timing constraints, their cascade yields an end-to-end reliability guarantee at the system level.
% : by exploiting the micro-architectural characteristics of each key module and the mathematical properties of its internal operators (e.g., the linearity of matrix operations in the systolic array), we tailor dedicated safeguards to each module and insert lightweight reliability operators at module boundaries, thereby providing end-to-end protection with minimal system-level overhead. 
% Under this decomposition, all intermediate states and outputs necessarily pass through protected modules and their boundary checks; as long as each module offers explicit, bounded reliability guarantees within its timing constraints, the cascade of these modules naturally yields an end-to-end reliability guarantee for the whole NPU.
% For critical registers,\zhenote{This is the first time to mention ``critical registers''. Can we have an explicit definition of them?} which feature fixed bit-width and high access frequency, we employ lightweight Single Error Correction and Double Error Detection (SEC-DED) ECC to enable protection, covering instruction-flow state (e.g., reservation station and control units, Fig.~\ref{overview}.\blacka\ and \blackb\ ) and constants for non-linear operators passed from software APIs (e.g., $ln2$ used by ReLU).
For critical registers on the instruction and control paths, such as issue/scheduling registers (e.g., Fig.~\ref{overview}.\blacka\ and \blackb\ ) and constants injected via software APIs (e.g., $ln2$ used by ReLU), the state size is small but the fan-out is large, making them sensitive to faults while also requiring low access latency. We therefore employ lightweight Single Error Correction and Double Error Detection (SEC-DED) ECC to provide protection with tightly bounded overhead.
% In contrast, local memory, which involves high-throughput data movement, is safeguarded via a checksum scheme: a guardpad stores checksums (Fig.~\ref{overview}.\blackg\ ) and a guardlinker binds data to checksums and logs fault locations (Fig.~\ref{overview}.\blacke\ ).\zhenote{OK, we mentioned that the local memory is different from the registers that involve high-throughput data movement. So, what is the key feature of teh guardpad and guardlinker to satisfy such requirement? Why ECC does not work?}
In contrast, local memory involves high-throughput, highly parallel data movement.
Naively applying strong ECC would require wider check bits and more complex decoders, offering limited capability for multi-bit error detecting.
We therefore design a checksum-based guard mechanism: the guardpad stores checksums (Fig.~\ref{overview}.\blackg\ ) and the guardlinker binds data to checksums and logs fault locations (Fig.~\ref{overview}.\blacke\ ).
This decouples checking logic from the local memory and pipelines checksum generation and verification at block granularity along the high-bandwidth access path, better matching the local-memory dataflow and enabling efficient detection and correction of multi-bit errors.
% The systolic array, with high computational density, data reuse and highly sensitive to execution rhythm disruptions, is checked by a hardware-redesigned, ABFT-inspired shield group (Fig.~\ref{overview}.\blackk\ ) running in parallel with the array (Fig.~\ref{overview}.\blackh\ ) via a configuration strategy and a pipelined design (Sec.~\ref{sec:hardware.3}), yielding negligible slowdown.
% \zhenote{Why does the ABFT work? What is the difference between our ABFT and the existing ones.}
For the systolic array, its high compute density make it highly sensitive to execution rhythm; any in-array re-execution or serial checking can easily disrupt the pipeline.
Leveraging the linearity of the matrix MAC operation, we introduce a hardware-redesigned, Algorithm-Based Fault Tolerance (ABFT)-inspired shield group (Fig.~\ref{overview}.\blackk\ ).
Through a fully decoupled design, offline configuration, and pipelined optimisation (Sec.~\ref{sec:hardware.3}), the shield group runs in parallel with the array, yielding negligible slowdown.
% Non-linear operators often have distinct properties and low overhead. Those with clear properties can be checked directly, while others are protected using TMR.
For non-linear operators (Fig.~\ref{overview}.\blackc\ ), which often have distinct properties and low overhead, we combine property-based checks when algebraic invariants are explicit and TMR when they are not.
We also design some ISA extensions that bridge the hardware mechanisms with the software stack, enabling the runtime to observe protection status and perform targeted pipeline preprocessing.

\noindent \textbf{Framework workflow.} 
In NPUs, software APIs split the input into blocks and load them row‑by‑row into the scratchpad via multiple \textsl{mvin} instructions (Fig.~\ref{overview}.\blackf\ ). 
While the DMA streams each matrix, a checksum adder (Fig.\ref{overview}.\blackd\ ) computes all row/column checksums in a single pass (detailed in Sec.~\ref{sec:hardware}.B), and stores them in the guardpad and guardlinker (Fig.~\ref{overview}.\blacke\ and \blackg\ ).
When a \textsl{compute} instruction is dispatched to the NPU, it generates a \textsl{precompute} sub-instruction, which is issued alongside the \textsl{compute} instruction (Fig.~\ref{overview}.\blacka\ ). 
The \textsl{precompute} pre-schedules the verifier and corrector   (Fig.~\ref{overview}.\blackd\ ) to detect and correct the original data block.
Given that the incoming matrix may only partially cover the original block, the checksum adder subsequently appends checksums to the incoming matrix after error correcting, enabling verification during subsequent computations.
During the multiplication, we intercept the input stream to the systolic array and direct the column checksums of the weight matrix, the full input matrix, the row checksums of the input matrix, and the transposed weight matrix
into the shield group (Fig.~\ref{overview}.\blackk\ ) for verification.
Verified results and their checksums are written back to the accumulator and guardpad, and the write‑back phase repeats checksum verification and correction.
% All stages — load, compute, write‑back — run in a pipeline to minimise slowdown.
Furthermore, based on the conclusions in Sec.~\ref{sec:motivation}, low-impact bit positions (e.g., mantissa) are ignored to reduce cost.
In addition, to aid permanent fault localisation, the error block in the guardlinker (Fig.\ref{overview}.\blacke\ ) records faulty locations (e.g., memory addresses and tiles).
Upon tasks completion, Strix issues the instruction \textsl{mvout\_error\_block} to report the affected block.
% , enabling developers to identify permanent faults.
% At the system level, we also design some ISA extensions to facilitate the deployment, as summarised in Tab.~\ref{isa}.
% \zhenote{Two problems: 1) the table 1 is referred too late, and we have described some of these instructions in the above sentences.
% 2) Some of the ISAs have never been used, e.g., flush, strix\_on/off. }

% \noindent \textbf{Supporting other NPUs.}
% Strix introduces only light design invasion to NPUs.
% Specifically, we only make incremental modifications to NPUs without disrupting the original data flow (i.e., no changes are required to the programming model, DNN algorithms, workloads, or APIs). 
% This allows Strix to be adapted to other NPUs based on systolic arrays and local memory~\cite{qin2020sigma,shin2018dnpu,chen2019eyeriss,amert2017gpu}.
% whilst preserving compatibility with diverse scheduling strategies, such as software API-based scheduling~\cite{genc2021gemmini} and CISC-driven scheduling~\cite{genc2021gemmini,chen2019eyeriss}.

% \zhenote{I remember that reviewers attacked this subsection. We have given the reason for using the Gemmini, can we delete this subsection directly?}

\vspace{-10pt}
\section{The Micro-architecture of Strix}
\label{sec:hardware}

% Having established the system-level design principles and protection strategy in Sec.~\ref{sec:approach}, we now describe the concrete micro-architectural realisation of Strix on the NPU.
% As introduced in Sec.~\ref{sec:approach}, we design a new micro-architecture for Stirx, consisting of: 
% (i) SEC-DED ECC for critical registers; 
% (ii) a guardlinker and guardpad to verify local memory; 
% (iii) a shield group for parallel checksum computation; and 
% (iv) operator-level protections for non-linear operators based on their properties.

\subsection{Reliability of Registers}
Registers in NPUs are usually responsible for storing instructions, constants, and other information.
% The reliability of registers is directly tied to the correct execution, as erroneous decoding can lead to computation failures or program crashes~\cite{ibrahim2020soft,mittal2020survey}.
The fixed bit-width and frequent access of registers make ECC a cost-effective strategy.

\begin{figure}[t]
\centering
\includegraphics[width=0.6\linewidth]{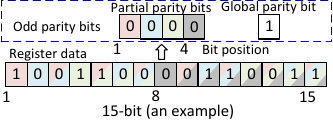}
\vspace{-10pt}
\caption{SEC-DED ECC: each partial parity bit verifies data of the same colour, and the global parity bit verifies the partial parity bits.}
\label{ecc}
\vspace{-14pt}
\end{figure}

\noindent \textbf{Protection mechanism.} 
Fig.~\ref{ecc} generates odd parity by first computing a global parity bit and then deriving partial parity bits from data subsets indexed by bit positions. 
Each partial parity bit monitors the original data where specific binary indices are set to 1. For example, in Fig.~\ref{ecc}, the red partial parity bit (binary 001) is generated from the subset of the original data where the last binary index is set to 1 (i.e., the red bit of the register, binary 0001 etc.). 
% When an error occurs, deviations in the expected parity values enable fault localisation. 
If both the red and blue partial parity and the global parity mismatch, the fault can be located at bit 0011. 
Conversely, if partial parities mismatch but the global parity is correct, a multi‑bit error is inferred.
The mechanism introducing a cost of $(\lceil \log_2 (\alpha + 1) \rceil + 1)$ bits ($\alpha$: the register width).

\subsection{Reliability of Local Memory}
% In NPUs, local memory stores both inputs and intermediates.
% However, errors in local memory can lead to widespread propagation in parallel computations, which may result in incorrect outputs across multiple subsequent layers, ultimately causing task failure~\cite{ibrahim2020soft}. 
In Gemmini, local memory, including scratchpad and accumulator, stores data in rows, utilising double-buffering for performance optimisation, with the amount of data per row aligned to the PEs in each row. 
Given the high-throughput characteristic of local memory, directly applying ECC is inadequate for multi-bit errors and would incur significant overhead.
Thus, we adopt a dual-vector checksum scheme to enable error detection and correction while avoiding intrusive modifications to the original design.
We designed a guardpad for storing checksums, and a guardlinker including a linker block and an error block.
The linker block ensures the mapping between data and checksums and logs faulty memory rows and tiles, tracking error times to assist developers in localising permanent faults.

\begin{figure}[t]
\centering
\includegraphics[width=1\linewidth]{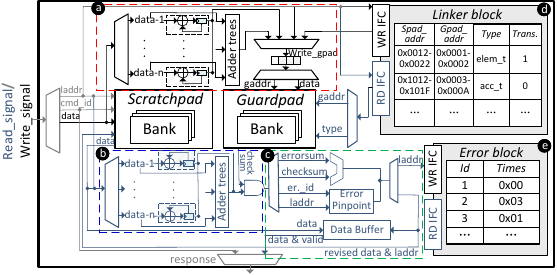}
\vspace{-20pt}
\caption{Components related to local memory reliability. Blue lines: read path; black lines: write path; grey lines: common path. 
% \zhenote{Better, but I don't understand the highlights of this figure.}
}
\label{mem}
\vspace{-14pt}
\end{figure}

\noindent \textbf{Protection mechanism.} 
Fig.~\ref{mem} shows the micro-architecture working in coordination within the local memory. 
Specifically, when DMA is writing data, the linker block (Fig.~\ref{mem}.\blackd\ ) and checksum adder (Fig.~\ref{mem}.\blacka\ ) tap the stream. The checksum adder is paired with a set of adder trees to generate the row checksums, and a group of adder-register units to generate the column checksums, which are then written into the guardpad.
During data reading, the data verifier (Fig.~\ref{mem}.\blackb\ ) fetches checksums for verification. 
If a mismatch appears, the data corrector (Fig.~\ref{mem}.\blackc\ ) uses the error information, including row and column positions and the discrepancy with the checksum, to locate the fault, correct the data, and update the error block (Fig~.\ref{mem}.\blacke\ ). 
The corrected data is then forwarded from the buffer within the corrector.
Less‑critical bits (e.g., mantissas in our study) are optionally ignored to cut cost; more generally, the bit-selection policy is developer-tunable to accommodate other models.

\noindent \textbf{Adder tree design.} 
In local memory, checksums serve solely to verify the correctness of the original data and never participate in the matrix multiplication. Thus, the adder trees treat the data as binary streams and perform binary addition. For example, adding 1100 and 0011 always yields 1111, with no FP conversion required.
Overflow is likewise benign.
% : if it occurs during accumulation, it will be reproduced during the verification step.
Importantly, this is limited to local memory.
% ; the systolic array deploys pipelined full‑precision guard operators, sustaining throughput with low latency.

\noindent \textbf{Error detection and correction.} 
If the recomputed and stored checksums diverge, we first locate the intersection of the mismatching row and column. With a single mismatch, the error is uniquely correctable by the checksum delta.
However, when multiple row and column mismatches occur, ambiguity in fault localisation arises. 
Thus, we exploit the property that a single erroneous data induces consistent discrepancies between the computed sum and the checksum across both its row and column. 
It enables cross-localisation, allowing for error correction based on the observed discrepancies.
Finally, if only one of the two orthogonal checksums is invalid while the other remains correct, we attribute the fault to the checksum itself (not to the data), thereby preventing false‑positive corrections.

\subsection{Reliability of the Systolic Array}
\label{sec:hardware.3}
% The systolic array executes the matrix MAC kernels, which consume over 70\% of inference time~\cite{dos2023understanding}. 
% Faults introduced here quickly cascade through later layers, compounding their impact on overall model accuracy~\cite{ibrahim2020soft,mittal2020survey}.
In Gemmini, data fetching is controlled by configuration registers (e.g., stride). Thus, the fetched data may not always be the original block.
The computation modes supported by Gemmini include Output-Stationary Mode (OS-M) and Weight-Stationary Mode (WS-M).
Matrix operations follow $C = A \times B + D$, where $A$ is the input matrix, $B$ is the weight matrix, and $D$ is the bias matrix. In WS-M, $B$ is preloaded, while $A$ is fed in. The partial sums are stored in the accumulator and later added to $D$. In OS-M, $D$ is preloaded, while $A$ and $B$ are streamed.
Moreover, API‑defined sub‑matrix sizes must match the PE‑row width.
Thus, integrating fault-tolerant algorithms directly would need changes to the hardware architecture, software APIs, and workloads.
% Hence, we optimised the ABFT execution logic and deployed a shield group for checksum vector computations. 
Hence, unlike interleaved, detection-only schemes~\cite{xie2025realm}, we deploy a shield group fully decoupled from the systolic array. This decoupling preserves the array’s execution rhythm and produces row/column checksums to enable error correction.
The shield group achieves low-slowdown verification through: (i) a configuration strategy enabling parallel execution with the array, and (ii) a pipeline allowing concurrent operation with guard operators.

\begin{figure}[t]
\captionsetup[subfigure]{skip=2pt}  % 控制图与caption之间的垂直距离
\centering
\begin{subfigure}[t]{\linewidth}
    \centering
    \includegraphics[width=\linewidth]{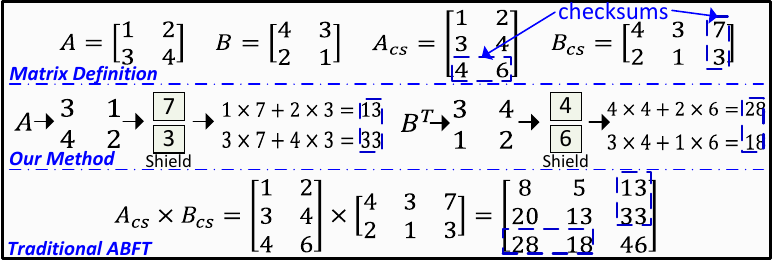}
    \caption{Mathematical principle and an example of the shield group.}
    \label{shielda}
\end{subfigure}

\begin{subfigure}[t]{\linewidth}
    \centering
    \includegraphics[width=\linewidth]{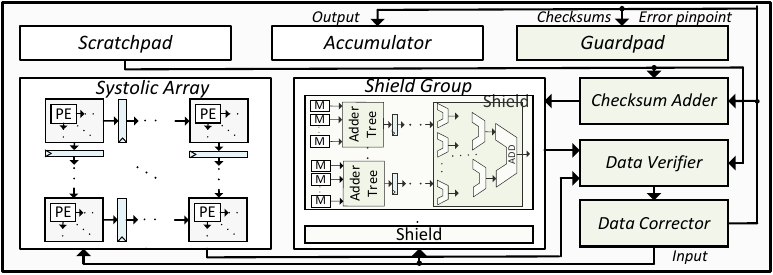}
    \caption{Protection mechanism of the shield group.}
    \label{shieldb}
\end{subfigure}

\begin{subfigure}[t]{\linewidth}
    \centering
    \includegraphics[width=\linewidth]{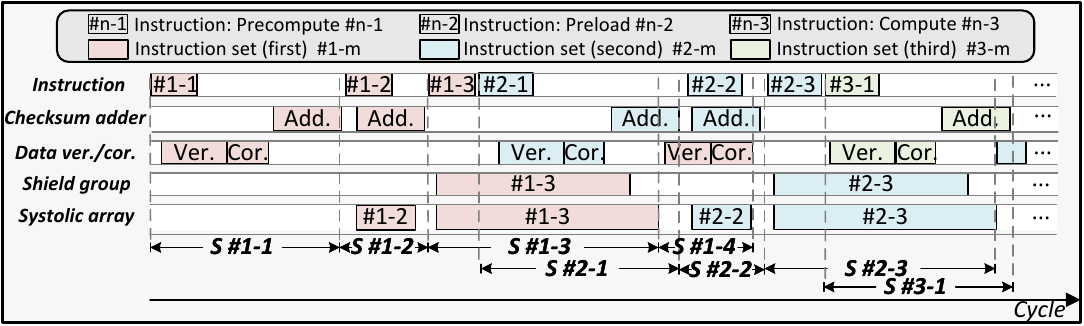}
    \caption{\npumodel\ pipeline in computational processes.}
    \label{shieldc}
\end{subfigure}

% \hspace{10pt}
\vspace{-10pt}
\caption{The shield group. In (c), $\#n-m$ denotes the m-th type of instruction in the n-th group, while $S \#n-m$ is different pipeline stages.}
\label{shield}
\vspace{-12pt}
\end{figure}

\noindent \textbf{Principle.} 
Fig.~\ref{shield}(\subref{shielda}) shows the principle of the shield group. We first add column checksums to $A$ and row checksums to $B$. Fixing $B$'s column checksums, we stream $A$ row by row. Each row is multiplied by its corresponding checksum and accumulated to compute the row checksum for $A\times B$. Similarly, fixing the row checksums of $A$ and streaming in the transpose of $B$ ($B^T$ in Fig.~\ref{shield}(\subref{shielda})), we compute the final column checksums. Finally, the results are identical to ABFT.

\noindent \textbf{Protection mechanism.} 
Figs.~\ref{shield}(\subref{shieldb}) and \ref{shield}(\subref{shieldc}) illustrate the shield group’s micro-architecture and its pipeline. Specifically, the shield employs a set of multipliers aligned with the number of PEs per row and several small adder trees. 
When a \textsl{compute} instruction is dispatched to the NPU, the system concurrently generates a \textsl{precompute} sub-instruction, which is issued alongside the corresponding \textsl{compute}, establishing explicit dependencies with both \textsl{preload} and \textsl{compute}.
In WS-M, the \textsl{precompute} executes first, triggering the data verifier and corrector to ensure original data integrity. The checksum adder then computes checksums for $A$ (stage \textit{S \#1-1} in Fig.~\ref{shield}(\subref{shieldc})).
While $B$ is preloaded, the transposer and checksum adder receive the data stream, performing transposition and checksum computation (\textit{S \#1-2} in Fig.~\ref{shield}(\subref{shieldc})). 
At this stage, the shield group has acquired all necessary elements.
Once the \textsl{compute} arrives, the shield group executes in parallel with the systolic array (\textit{S \#1-3} in Fig.~\ref{shield}(\subref{shieldc})). Upon completion, the data verifier compares the results, while the data corrector performs fault localisation, error correction (\textit{S \#1-4} in Fig.~\ref{shield}(\subref{shieldc})). 
Finally, the result is added to $D$, and similarly, in the guardpad, we yield the final checksums.
In OS‑M, the flow is identical, except checksums of $D$ are generated when its preload.

\noindent \textbf{Pipeline design.} 
Strix masks guard operator latency with a four‑stage pipeline (see Fig.~\ref{shield}(\subref{shieldc})).
The latency introduced by Strix manifests in the stages 1 and 4.
To conceal these delays, we leverage pre-execution of \textsl{precompute}, ensuring that the first stage overlaps with the third stage (computing) and the fourth stage overlaps with the second stage (data preloading). 
For instance, in Fig.~\ref{shield}(\subref{shieldc}), \textit{S \#2-1} is overlapped by \textit{S \#1-3}, and \textit{S \#1-4} is overlapped by \textit{S \#2-2}.
As a result, this design enables near-zero-latency data processing by guard operators.

\noindent \textbf{Shield configuration.} 
To hide shield-induced latency (WS-M shown; other modes analogous), we size the shield group so that its total latency $\sigma$ does not exceed the array’s matrix window $L_{SA}$. Let $I$ be the tiles per row of the array, $J$ the PEs per row within a tile, and $K$ the number of shields. 
For an $I\cdot J \times I\cdot J$ matrix, $L_{SA}$ is the interval from the first input’s arrival to the last output’s departure: the final input arrives after $I\cdot J+I-1$ cycles and needs a further $I$ cycles to traverse the array, hence $L_{SA}=I\cdot J+2\cdot I-1$ cycles~\cite{gemmini_systolic_transposer}.
Each shield ingests one vector per cycle and uses multiplier rows aligned with array rows plus adder trees whose first level fan-in is at most $2^{J-1}$. The resulting shield latency is: 
\begin{equation}
\scriptsize
\sigma = \frac{2 \cdot I \cdot J}{K} + 1 +   \left(\max \left\{ 0, \left\lfloor \frac{1}{J} \cdot \log_2 \left\lceil \frac{I \cdot J}{2^{J}} \right\rceil \right\rfloor \right\} + 1\right), 
\end{equation}
where $\frac{2 \cdot I \cdot J}{K}$ accounts for processing the two matrices across $K$ shields, and the last two terms capture pipeline start-up/adder-tree depth. Enforcing $\sigma \leq L_{SA}$ yields the minimum shield parallelism:
\begin{equation}
\scriptsize
K =\left\lceil \frac{2 \cdot I \cdot J}{I \cdot J + 2 \cdot I - 3 - \max \left\{ 0, \left\lfloor \frac{1}{J} \cdot \log_2 \left\lceil \frac{I \cdot J}{2^{J}} \right\rceil \right\rfloor \right\} } \right\rceil.
\label{13}
\end{equation}
By configuring $K$ shields, we can overlap shield latency with array execution, reducing the slowdown incurred by reliability deployment.

\begin{figure}[t]
\centering
\includegraphics[width=\linewidth]{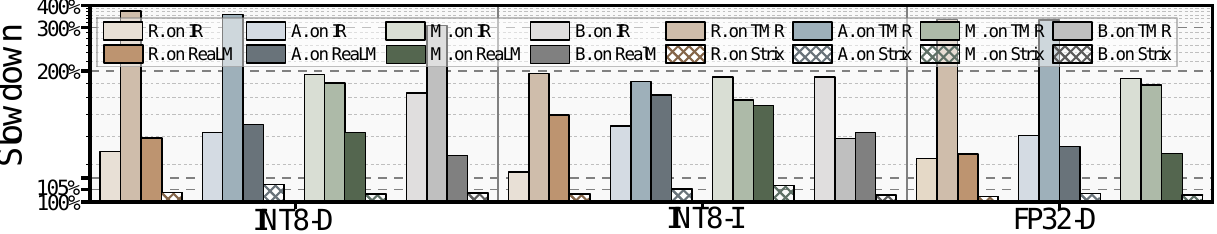}
\vspace{-22pt}
\caption{Performance overhead (R.: ResNet-50; A.: AlexNet; M.: MobileNet-V2; B.: BERT). Strix achieves a geometric of $1.04\times$.}
\label{slowdown}
\vspace{-12pt}
\end{figure}

\begin{figure}[t]
\captionsetup[subfigure]{skip=2pt}  % 控制图与caption之间的垂直距离
\centering
\begin{subfigure}[t]{\linewidth}
    \centering
    \includegraphics[width=\linewidth]{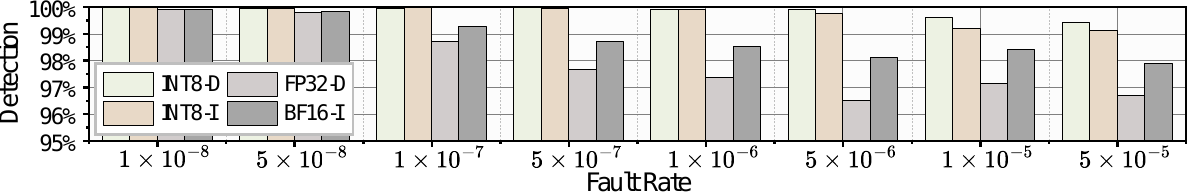}
    \caption{Error detection coverage of Strix.}
    \label{fig:cover1}
\end{subfigure}

\begin{subfigure}[t]{\linewidth}
    \centering
    \includegraphics[width=\linewidth]{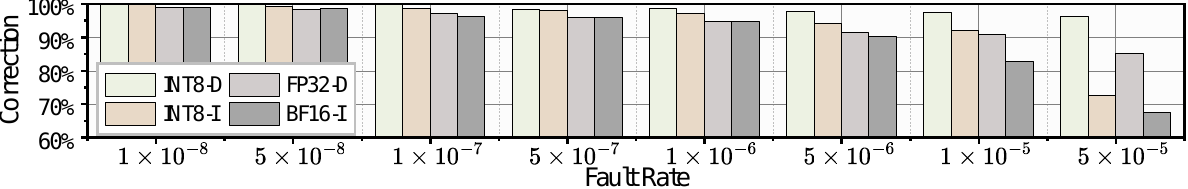}
    \caption{Error correction coverage of Strix.}
    \label{fig:cover2}
\end{subfigure}
\vspace{-12pt}
\caption{The error detection and correction coverage of Strix.}
\label{cover}
\vspace{-14pt}
\end{figure}

\noindent \textbf{Adder tree design.} 
Deep adder trees can affect circuit timing closure. 
To mitigate this problem, and considering that each PE within a tile consists of combinational logic and a MAC design, we cap their depth to the PEs‑per‑row of each tile — matching the NPU’s native critical path instead of extending it. 
% This allows the critical path to be determined by the NPU's inherent configuration, rather than Strix.
We also prune the redundant adders in the edge adder trees to improve resource utilisation.
% \zhenote{No discussions regarding the differences between the adder tree here and the ones in the memory protection. Can we fix?}  

\noindent \textbf{Data overflow.} 
% Unlike the data stored in local memory, overflow in computed data can compromise verification, necessitating preventive measures. 
Computed values, unlike static data in local memory, can overflow and break verification.
For original data, PEs incorporate overflow prevention mechanisms, employing bit expansion strategies during MAC. 
% Furthermore, the output data format typically has a higher bit width. 
% For instance, INT8 inputs are typically computed and stored in INT32 format.
% As a result, overflow is highly unlikely within the PE’s internal computations.
We also ensure that the checksum uses the same bit width as the output to provide sufficient margin against overflow.

\noindent \textbf{Fault localisation.} 
% In WS-M, an error occurring within a tile typically affects the entire column and potentially subsequent columns~\cite{agarwal2023towards}. When consecutive checksum mismatches are detected, the system can infer the presence of a faulty tile and localise it to the earliest column where discrepancies appear. In OS-M, due to the lack of clear data dependencies~\cite{agarwal2023towards}, the faulty tile can be accurately identified by locating the earliest row and column with checksum mismatches.
In WS‑M, a tile fault corrupts its own column and often later ones~\cite{agarwal2023towards}; consecutive checksum mismatches allow Strix to infer a faulty tile and localise it to the earliest column with discrepancies.
% When consecutive checksum mismatches, Strix can infer the presence of a faulty tile and localise it to the earliest column with discrepancies.
In OS-M, the faulty tile can be accurately identified by locating the earliest row and column with mismatches~\cite{agarwal2023towards}.

\subsection{Reliability of Non-linear Operators}
The overhead of non-linear operators in NPUs hinges on their implementation. For instance, ASIC units are accurate but expensive~\cite{talpes2022dojo}, whereas approximate designs are lightweight~\cite{genc2021gemmini}.  
We therefore verify the former via operator‑specific mathematical invariants, and the latter with a mix of invariants and redundancy. 
In Gemmini, non-linear operators include LayerNorm, ReLU, Softmax, GELU, and Pooling, with each contributes less than 1\%  overhead~\cite{genc2021gemmini}. 
Therefore, for operators that have clear algebraic properties, we verify based on these properties. Conversely, for operators lacking properties or whose rule is complex, we ensure reliability through redundancy.

Specifically, for LayerNorm, we exploit the sum of the non-affine normalised activations should be 0 to enable a consistency check.
For Softmax, we perform a quick verification based on its output sum should be 1.
In contrast, ReLU, GELU, and Pooling also exhibit certain properties but require more complex verification. Thus, we adopt a redundant copy approach. As a result, we reduce the verification overhead while maintaining high coverage for functional correctness.

\section{Experimental Evaluation}
\label{sc:EE}

\subsection{Experimental Setup}

\noindent \textbf{Platform.}
% We deploy Strix on an FPGA cluster with 20 AMD Virtex UltraScale+ VU19P FPGAs.
% We configure four setups to evaluate the performance of Strix under three widely adopted data formats~\cite{dettmers2022gpt3,johnson2018rethinking,ahn2023performance,zhou2024survey}:
We deploy Strix on a 20-node FPGA cluster (AMD Virtex UltraScale+ VU19P) with four setups:
(i) INT8 default (INT8-D): based on the default open-source Gemmini, including a 256KB scratchpad, 64KB accumulator, and a $16 \times 16$ systolic array (one PE per tile);
(ii) INT8 industrial (INT8-I): 2048KB scratchpad, 512KB accumulator, $128 \times 128$ systolic array ($4 \times 4$ PEs per tile, same as TPU v3~\cite{jouppi2020domain});
(iii) FP32 default (FP32-D): 1024KB scratchpad, 64KB accumulator, $16 \times 16$ systolic array (one PE per tile);
(iv) BF16 Industrial (BF16-I): 4096KB scratchpad, 1024KB accumulator, $128 \times 128$ systolic array ($4 \times 4$ PEs per tile).
Due to FPGA resource limits, BF16-I is simulated in PyTorch following the Spike-modelled behaviour~\cite{riscv-spike} and used only for LLM performance analysis.
% The software stack is compiled using the RISC-V GNU toolchain.
% The system runs at 100 MHz in practice, but a target environment of 500 MHz was emulated by integrating the FASED memory timing model.

\begin{figure}[t]
\captionsetup[subfigure]{skip=2pt}  % 控制图与caption之间的垂直距离
\centering
\begin{subfigure}[t]{\linewidth}
    \centering
    \includegraphics[width=\linewidth]{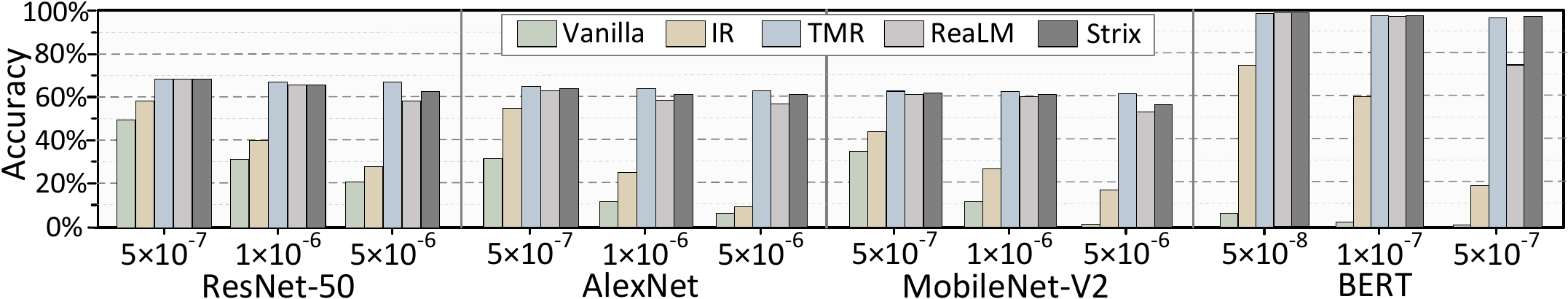}
    \caption{Impact of different strategies on DNN accuracy under INT8-D.}
    \label{accu1}
\end{subfigure}

\begin{subfigure}[t]{\linewidth}
    \centering
    \includegraphics[width=\linewidth]{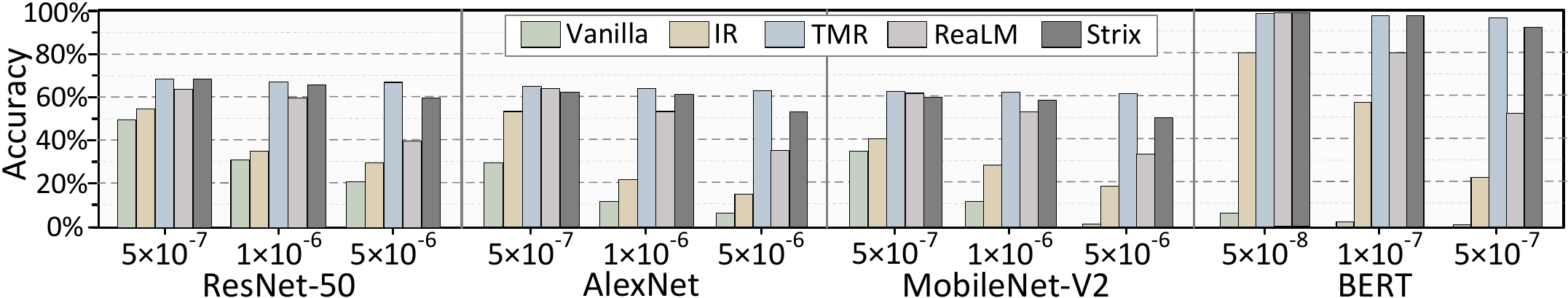}
    \caption{Impact of different strategies on DNN accuracy under INT8-I.}
    \label{accu2}
\end{subfigure}

\begin{subfigure}[t]{\linewidth}
    \centering
    \includegraphics[width=\linewidth]{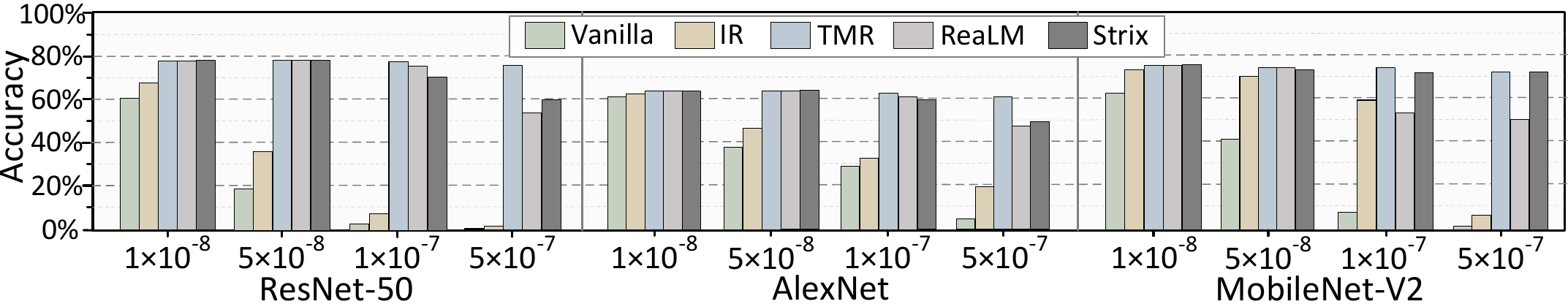}
    \caption{Impact of different strategies on DNN accuracy under FP32-D.}
    \label{accu3}
\end{subfigure}
\vspace{-12pt}
\caption{Impact of different strategies on DNN accuracy across various configurations and fault rates. Vanilla denotes the unprotected system.
}
\label{accu}
\vspace{-14pt}
\end{figure}

\noindent \textbf{Models and datasets.}
To evaluate Strix, we benchmark both DNNs and LLMs.
The DNN suite covers three mainstream families: CNNs (AlexNet~\cite{krizhevsky2012imagenet}, MobileNet-V2~\cite{sandler2018mobilenetv2}), residual networks (ResNet-50~\cite{he2016deep}), and transformer-based models (BERT~\cite{kenton2019bert}). Owing to Gemmini limitations, BERT is evaluated at INT8, whereas the remaining DNNs are tested at both FP32 and INT8.
For LLMs, we choose three widely used models: Llama‑3.2‑1B~\cite{dubey2024llama}, Qwen‑2.5‑3B~\cite{hui2024qwen2}, and Mistral‑7B~\cite{Mistral7B}.
For datasets, image classification models are evaluated on the Tiny ImageNet~\cite{yao2015tiny}, BERT is evaluated on the DBPedia-14~\cite{rangwani2022cost}, and the LLMs are evaluated on the LAMBADA~\cite{paperno2016lambada}. 
For all cases, the first 1,000 samples are used for evaluation.

\noindent \textbf{Fault injection.}
We implement multiple types of fault injection within the hardware of \npumodel. The injection targets include original data, checksums, and hardware components. The injection rate is defined relative to the bit count of the original input.
We inject both transient
and permanent faults via custom ISA extensions.
Transient faults are injected by randomly flipping bits at each pipeline stage during execution. 
Permanent faults are injected following the method proposed in~\cite{agarwal2023resilience}, where additional hardware logic is incorporated to enable custom instructions that forcibly fix selected signals to 1 or 0.

\noindent \textbf{Baseline setup.}
The default fault injection rate is $1\times 10^{-7}$.
% For comparison, we define three baseline groups representing common hardware and software fault tolerance schemes.
% We compare Strix against three baseline groups that represent (i) hardware scheme, (ii) software scheme, and (iii) SOTA detect-and-rollback scheme—each implemented under a hardware budget area-matched to Strix for a fair comparison.
We compare against three baselines representing common hardware/software schemes and a contemporary SOTA, all implemented under a hardware budget area-matched to Strix for a fair comparison:
(i) Hardware (partial TMR). Registers and local memory are protected with ECC, while the systolic array employs TMR.
(ii) Software (triple IR). Sensitivity-guided redundancy: early layers for CNNs (AlexNet first two convolutions; first convolution of each MobileNet-V2 block; first ResNet-50 residual block~\cite{guo2023neural,ibrahim2020soft}); feed-forward layers for BERT~\cite{gao2025dependability}; and the first 20\% / last 10\% of layers for LLMs~\cite{he2025fine}.
(iii) SOTA (ReaLM~\cite{xie2025realm}). Online detection with selective rollback; to avoid unbounded latency under permanent faults, at most one rollback attempt is permitted.
% The hardware baseline employs the partial TMR, designed to match Strix’s hardware overhead. Specifically, registers and local memory are protected using ECC, while the systolic array uses TMR.
% The software baseline applies triple IR based on the fault sensitivity of each network. Specifically, for CNNs, where early convolutional layers are known to be more fault-sensitive~\cite{guo2023neural}, we apply redundancy to the first two convolutional layers of AlexNet and the first convolution of each block in MobileNet-V2. For ResNet-50, redundancy is applied to its first residual block~\cite{ibrahim2020soft}. For BERT, redundancy is applied to the feed-forward layers~\cite{gao2025dependability}.
% For LLMs, redundancy is applied to layers near the input and output (i.e., the first 20\% and last 10\%)~\cite{he2025fine}.
All metrics are averaged over 10 runs.

% \vspace{-12pt}
\subsection{Observations and Results Analysis}

% \subsection{Performance Overhead}

\noindent \textbf{(1) Performance overhead.}
Fig.~\ref{slowdown} presents the performance overhead across different schemes.
% Overall, conventional mechanisms exhibit significant sensitivity to both computational resource usage and execution efficiency. 
% For instance, TMR consistently demonstrates the most substantial performance overhead across all models and configurations, frequently exceeding 180\% and reaching up to 368\% in certain scenarios (e.g., ResNet-50 under INT8-D).
Overall, TMR incurs the highest slowdown, typically exceeding $1.80\times$, and reaching $3.68\times$ for ResNet-50 under INT8-D.
In contrast, IR shows greater variability ($1.13\times$ to $1.95\times$), indicating strong dependence on model structure and compute density.
ReaLM falls between $1.22\times$ to $1.59\times$; slowdowns grow with larger systolic arrays (e.g., INT8-D) because, at a fixed fault rate, per-matrix error incidence rises and each rollback carries higher cost.
By contrast, Strix maintains the lowest performance overhead across all evaluated scenarios, with slowdowns ranging from $1.02\times$ to $1.07\times$ and a geometric slowdown of $1.04\times$. 
This demonstrates that Strix achieves significantly better adaptability and scalability.

\begin{figure}[t]
% \vspace{-5pt}
\captionsetup[subfigure]{skip=2pt}  % 控制图与caption之间的垂直距离
\centering
\begin{subfigure}[t]{\linewidth}
    \centering
    \includegraphics[width=\linewidth]{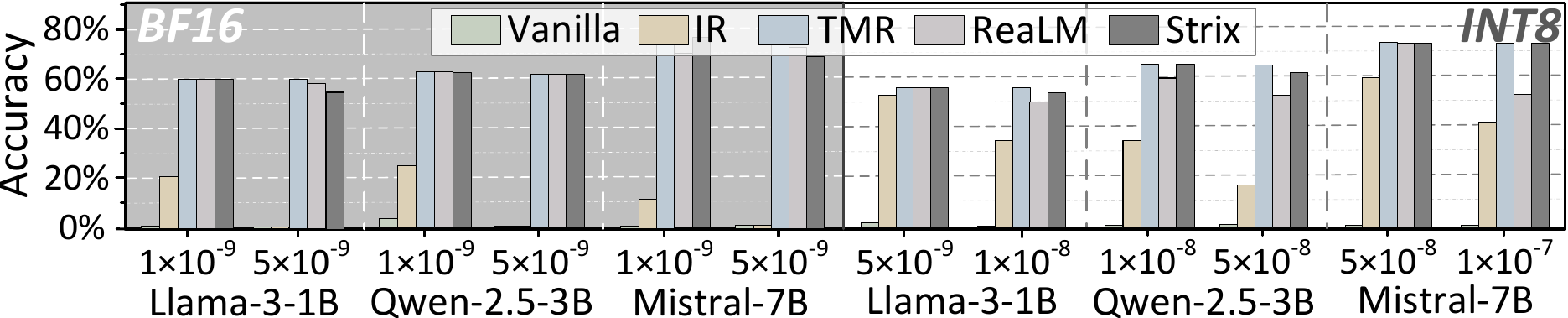}
    \caption{Impact of different strategies on LLM accuracy.}
    \label{llm1}
\end{subfigure}

\begin{subfigure}[t]{\linewidth}
    \centering
    \includegraphics[width=\linewidth]{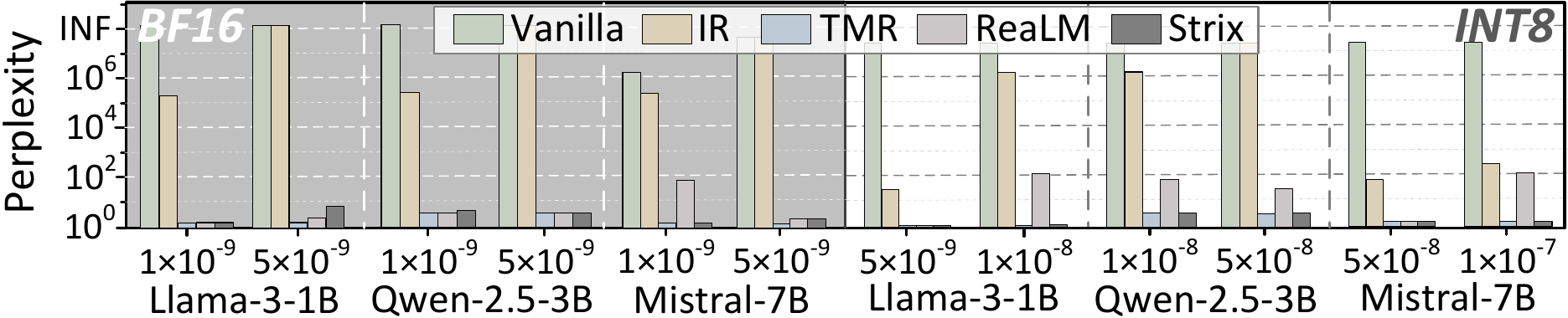}
    \caption{Impact of different strategies on LLM perplexity.}
    \label{llm2}
\end{subfigure}
\vspace{-12pt}
\caption{Impact of different strategies on LLM perplexity and accuracy, with the left half corresponding to BF16-I and the right half to INT8-I.
}
\label{LLMaccu}
\vspace{-15pt}
\end{figure}

% \subsection{Reliability Coverage}
% \noindent \textbf{Experimental setup.}
% We adopt the same settings used in Sec.~\ref{sc:EE}.A to assess the error detection and correction coverage of Strix under varying fault rates.

\noindent \textbf{(2) Detection coverage.}
Fig.~\ref{cover}(\subref{fig:cover1}) reports the error detection coverage of Strix. Overall, Strix exhibits consistently high capability in INT8-D and INT8-I, with coverage exceeding 99\%.
In FP32-D and BF16-I, Strix achieves over 97\% detection coverage under non-extreme fault rates (i.e., $\leq 10^{-6}$). The slight degradation at higher fault rates is attributable to:
(i) precision loss inherent to FP in hardware, and
(ii) faults injected into checksum bits, which can impair detection.

\noindent \textbf{(3) Correction coverage.}
Fig.~\ref{cover}(\subref{fig:cover2}) presents the correction coverage of Strix.
Under typical fault rates (i.e., $\leq 10^{-6}$), all configurations maintain a correction coverage exceeding 95\%.
However, as the fault rate increases, a noticeable decline is observed. Among the four configurations, INT8-I and BF16-I are most affected.
This trend primarily stems from the use of larger systolic arrays in these configurations. Under high fault rates, the increased number of involved processing elements raises the probability of multi-bit faults, which compromises the effectiveness of the correction mechanism.

% \subsection{LLM \& DNN Performance}

\noindent \textbf{(4) DNN performance.}
Fig.~\ref{accu} shows the variation in DNN accuracy across different settings. 
Compared to Vanilla, all strategies improve inference accuracy, but their effectiveness diverges as fault rates increase.
The IR offers moderate protection at low fault rates but degrades rapidly. 
ReaLM’s rollback mechanism fails under permanent faults; consequently, under higher fault rates, especially with larger systolic arrays and critical faults, model quality collapses, a trend that persists even after averaging across trials.
TMR behaves as expected, sustaining the highest accuracy across all rates.
In contrast, Strix mitigates accuracy loss within practical fault regimes, outperforming IR and matching ReaLM/TMR at low fault rates; only at very high rates does Strix show mild degradation, yet it remains better than ReaLM.
Across configurations, Strix yields larger gains in INT8-D and FP32-D. Although INT8-I poses greater challenges due to the larger array, Strix remains close to TMR at several fault settings.

\noindent \textbf{(5) LLM performance.}
Fig.~\ref{LLMaccu} shows the LLMs' performance versus fault rate. The overall trend mirrors the DNN results.
% As the fault rate rises, Vanilla quickly collapses.
All protections confer benefits, but IR provides only limited improvement at low rates.
ReaLM, again hindered by its inability to handle permanent faults, trails Strix, though the inherently lower fault tolerance of LLMs slightly narrows the gap.
By contrast, Strix maintains accuracy and perplexity nearly indistinguishable from TMR, alleviating performance degradation of LLMs in faulty environments.

% \subsection{The Worst-Case Detection Latency}
% % \noindent \textbf{Experimental setup.}
% We measure the worst-case error detection latency of each component within Strix under a frequency of 500MHz.
% To ensure the latency is incurred by Strix, we define detection latency as the time interval between the execution of the instruction and the moment it is detected.

\noindent \textbf{(6) The worst-case detection latency.}
We measure the worst-case error detection latency of each component within Strix under a frequency of 500MHz.
To ensure the latency is incurred by Strix, we define detection latency as the time interval between the execution of the instruction and the moment it is detected.
As shown in Tab.~\ref{latency}, the detection latency of Strix remains within the sub-micro-second. The higher latency observed in the systolic array is due to Strix's post-verification strategy, where verification is deferred until after the matrix computation and checksum generation are completed. 
% Thus, making the latency sensitive to the execution time of the instruction. 
% In contrast, the latency in local memory arises from the large amount of data typically involved during write-back, which requires additional cycles.

% \subsection{Analysis of the Critical Path}
% % \noindent \textbf{Experimental setup.}
% We tested the maximum frequency of the Gemmini and components introduced by Strix with different PE configurations in each tile to determine the critical path. During RTL synthesis, we applied different clock constraints to specify the target frequency and performed multiple rounds of synthesis until the tool could not meet the frequency.

\noindent \textbf{(7) Analysis of the critical path.}
% As shown in Tab.~\ref{freq}, the components introduced by Strix exhibit a higher maximum frequency than Gemmini across all configurations. This result indicates that the determining factor for the critical path remains the original NPU, rather than Strix.
We tested the maximum frequency of the Gemmini and components introduced by Strix with different PE configurations in each tile to determine the critical path. During RTL synthesis, we applied different clock constraints to specify the target frequency and performed multiple rounds of synthesis until the tool could not meet the frequency.
As shown in Tab.~\ref{freq}, the components introduced by Strix exhibit a higher frequency than Gemmini across all configurations, whether INT8 or FP32, and regardless of the number of PEs in each tile. Say, the determining factor for the critical path remains the original NPU rather than Strix.

\begin{table}[t]
        \centering
        \caption{The worst-case detection latency of Strix-added components.}
        \vspace{-10pt}
        \includegraphics[width=\linewidth]{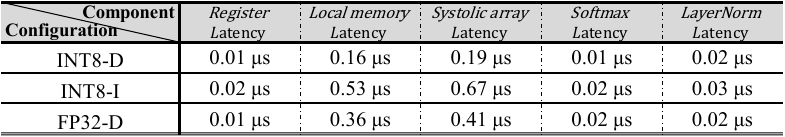}
        \label{latency}
        \vspace{-10pt}
        \caption{The max frequency of the Gemmini \& Strix-added components.}
        \vspace{-10pt}
        \includegraphics[width=\linewidth]{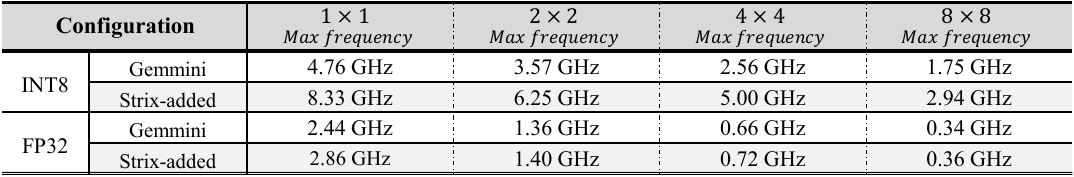}
        \label{freq}
        \vspace{-20pt}
\end{table}

% \subsection{Analysis of the Performance Gap for Shield Count}

% \noindent \textbf{Analysis of the performance gap for shield count.}
% Taking the ResNet-50 as an example, Tab.~\ref{shieldtest} shows the slowdown introduced by Strix under different shield counts, with values from Eq.~\ref{13} highlighted in grey. Increasing the number of shields yields diminishing returns beyond a certain point, which corresponds to the optimal upper bound for shield deployment. In each configuration, the analytically derived values are consistent with the measured optimal values, verifying the effectiveness of the analysis.

% \subsection{Hardware Overhead}
% % \noindent \textbf{Experimental setup.}
% We compare the \npumodel\ with Gemmini in terms of hardware overhead, and analyse the resource distribution across different components.
% % \footnote{The Guardpad size depends on local memory ($M$) and array dimension ($D$), as each $D \times D$ matrix needs  $2D$ checksums. With Gemmini’s double-buffering, it is approximated as $\frac{M}{2} \times \frac{2}{D}$.
% % Accounting for margins and omitting mantissa FP, all configurations use a 48KB Guardpad (32KB for scratchpad, 16KB for accumulator). The guardlinker is statically allocated with 8KB.}.
% % The synthesis of \npumodel\ is conducted using Vivado 2021.1 and Chipyard 1.10.0~\cite{amid2020chipyard}. Hardware overhead is primarily evaluated in terms of LUTs, registers, DSPs, RAMs, and power consumption.
% The hardware evaluation uses Synopsys Design Compiler (v2019.12) with the industry-standard TSMC 28nm technology library.

\noindent \textbf{(8) Hardware overhead.}
% We compare the \npumodel\ with Gemmini in terms of hardware overhead.
The hardware evaluation uses Synopsys Design Compiler (v2019.12) with the industry-standard TSMC 28nm technology library.
% \footnote{The Guardpad size depends on local memory ($M$) and array dimension ($D$), as each $D \times D$ matrix needs  $2D$ checksums. With Gemmini’s double-buffering, it is approximated as $\frac{M}{2} \times \frac{2}{D}$.
% Accounting for margins and omitting mantissa FP, all configurations use a 48KB Guardpad. The guardlinker is statically allocated with 8KB.}.
Across configurations, Strix adds 20.7\% area and 23.2\% power on INT8-D, 8.7\%/16.8\% on INT8-I, and 10.2\%/21.3\% on FP32-D. 
The higher overhead on INT8-D stems from its smaller baseline; INT8-I and FP32-D amortise the mechanisms more effectively.
The added area is dominated by the guardpad (13.9\%, 5.0\%, 4.6\% for INT8-D/INT8-I/FP32-D), followed by guard operators (3.4\%/3.5\%/5.3\%) as they are instantiated at pipeline boundaries to preserve continuous execution.
% , while the shield group is negligible (0.43\%/1.79\%/0.69\%)

\section{Conclusion}
In this paper, we present Strix, a full-stack framework for NPU reliability.
Strix demonstrates that effective and efficient protection must be grounded in a deep understanding of the system.
By re-partitioning the NPU from a reliability perspective, identifying key failure modes, and designing mitigation aligned with proposed insights, Strix has achieved near-complete fault coverage and near-zero detection latency, under minimal performance and area overhead. 
% In this paper, we introduce Strix, a generalised and full-stack framework for NPU reliability. Strix emphasises that effective and efficient protection strategies must be rooted in an in-depth understanding of the system and failure characteristics. 
% By re-partitioning the NPU from a reliability perspective, identifying critical failure modes, and aligning the mitigation strategies with the insights gained, Strix achieves near-complete fault coverage and sub-micro-second detection latency, under minimal performance and area overhead, offering a comprehensive solution for real-world deployment.

\section{Acknowledgement}
% We'd like to thank the reviewers for the helpful feedback. 
This work is supported by the National Key Research and Development Program (Grant No.2024YFB4405600), the U.S. National Science Foundation (Grant No.CNS-2340171), the Basic Research Program of Jiangsu (Grants No. BK20243042), and the Fundamental Research Funds for the Central Universities (No. 2242025K20013).

\balance
% \linespread{1}
\bibliographystyle{ACM-Reference-Format}
\bibliography{mybibfile}

\end{document}